\documentclass[conference]{IEEEtran}
\IEEEoverridecommandlockouts
\usepackage{cite}
\usepackage[ruled]{algorithm2e}
\usepackage{amsmath,amssymb,amsfonts}
\usepackage{algorithmic}
\usepackage{array}
\usepackage{balance}
\usepackage{cellspace}
\usepackage{graphicx}
\usepackage{hhline}
\usepackage{subfigure}
\usepackage{makecell}
\usepackage{multirow}
\usepackage{textcomp}
\usepackage[table]{xcolor}
\def\BibTeX{{\rm B\kern-.05em{\sc i\kern-.025em b}\kern-.08em
    T\kern-.1667em\lower.7ex\hbox{E}\kern-.125emX}}
\usepackage[left=0.625in, right=0.625in, top=0.75in, bottom=1.05in]{geometry}
\title{Enhanced Battery Degradation-Aware Scheduling for Distribution Network with Electric Vehicle Load\\
\thanks{This work was supported in part by A*STAR under its MTC Programmatic (Award M23L9b0052), MTC Individual Research Grants (IRG) (Award M23M6c0113), SIT’s Ignition Grant (STEM) (Grant ID: IG (S) 2/2023 – 792), and Future Communications Research \& Development Programme (FCP) under Grant FCP-SIT-TG-2022-007.}
\thanks{$^*$ Corresponding author.}
}

\makeatletter
\newcommand{\linebreakand}{%
  \end{@IEEEauthorhalign}
  \hfill\mbox{}\par
\mbox{}\hfill\begin{@IEEEauthorhalign}
}
\makeatother

\author{
\IEEEauthorblockN{Vijay Babu Pamshetti}
\IEEEauthorblockA{
\textit{Singapore Institute of Technology, Chaitanya Bharathi Institute of Technology}\\
vijaybabu.pamshetti@singaporetech.edu.sg, vijaybabup$\_$eee@cbit.ac.in}
\and
\IEEEauthorblockN{Wei Zhang$^*$}
\IEEEauthorblockA{
\textit{Singapore Institute of Technology}\\
wei.zhang@singaporetech.edu.sg}
\linebreakand
\IEEEauthorblockN{Andy, Man-Fai Ng}
\IEEEauthorblockA{
\textit{Agency for Science, Technology and Research}
\\ngmf@ihpc.a-star.edu.sg}
\and
\IEEEauthorblockN{Qingyu Yan}
\IEEEauthorblockA{
\textit{Nanyang Technological University}
\\alexyan@ntu.edu.sg}
\and
\IEEEauthorblockN{Kuan Tak Tan}
\IEEEauthorblockA{
\textit{Singapore Institute of Technology}
\\kuantak.tan@singaporetech.edu.sg
}
}

\begin{document}
\bstctlcite{IEEEexample:BSTcontrol}

\maketitle

\begin{abstract}
Batteries play a key role in today's power grid. In this paper, we investigate the impact of battery degradation on the distribution network. We formulate a multi-objective framework for optimizing battery scheduling with the goals of minimizing monetary costs and improving network performance. Our framework incorporates energy purchase and battery degradation into the costs and measures the network performance through energy losses and voltage deviation. We propose \texttt{Bach} for \underline{ba}ttery degradation-aware s\underline{ch}eduling based on $\varepsilon$-constraint and fuzzy logic methods. \texttt{Bach} is implemented for the IEEE 33-bus network for an experimental study. The results show the effectiveness of \texttt{Bach} in optimizing costs and performance simultaneously with battery degradation awareness and demonstrate the flexibility of further customization. 
\end{abstract}


\section{Introduction}
Solar photovoltaic (PV) systems and electric vehicles (EVs) have evolved beyond mere concepts for the smart grid. We are witnessing a rapid increase in PV integration into the power grid alongside the growing market share of EVs. These advancements not only address environmental concerns but also contribute significantly to sustainability. However, they introduce new challenges as the increased supply from PV and the heightened demand from EVs add substantial uncertainty to the grid. This uncertainty leads to increased energy losses and voltage deviations in the distribution network with a significant impact on the grid's quality of service \cite{chithra2022overview}.

An effective way to address the issues is the integration of battery energy storage systems (BESS) into distribution networks. Batteries can buffer surplus energy generated by PV systems and supply this stored energy during high demand periods, such as for EV charging \cite{lee2022novel} \cite{chai2021optimal}. This buffering concept is well-received by network operators for its effectiveness. However, this solution is not without drawbacks. The frequent charging and discharging cycles triggered to balance supply and demand dynamics accelerate battery degradation, and lead to high battery investment and potential sustainability issues.

Existing works have attempted to investigate and address the issues. A review of the challenges and the importance of battery health monitoring are presented in \cite{chithra2022overview}. The detailed cost modelling of battery degradation, where the depth of discharge (DoD) plays a major role, is developed in \cite{lee2022novel}. Battery degradation has been further incorporated into several scheduling algorithms for distribution networks. Examples include \cite{chai2021optimal} for distributed BESS, \cite{bahloul2024optimal} for BESS supporting multiple services, \cite{zhou2023novel,csensoy2023determination} for cooperative allocation of renewable energy, and \cite{cambambi2023model} incorporating time-varying prices. Despite these advancements, the focus is primarily on cost minimization for energy purchase and battery degradation, often neglecting the performance of distribution networks. This oversight is concerning, as optimizing cost often compromises network performance, or even violates the quality of service.

In this paper, we take network performance into account and propose the optimized \underline{ba}ttery degradation-aware s\underline{ch}eduling algorithms called \texttt{Bach}. We broaden the spectrum of the scheduling to capture realistic aspects of distribution networks with BESS, PV generation, EV load, and other loads. We develop a multi-objective framework to optimize both monetary costs (energy purchase and battery degradation) and network performance (energy loss and voltage deviation). To solve the formulated multi-objective problem, we propose two methods based on $\varepsilon$-constraint and fuzzy logic, respectively. An experimental study based on the IEEE 33-bus network is conducted to evaluate the impact of battery degradation on scheduling and parameter sensitivity. The study demonstrates the effectiveness of our proposed framework and methods. Besides the competitive performance, \texttt{Bach} also sheds lights on other applications like EVs with battery degradation issues. 

\section{Problem formulation}
We formulate a bi-objective problem of battery degradation-aware scheduling, considering monetary cost and the performance of distribution networks.

\subsection{Objective 1: Monetary Cost}
The monetary cost includes the cost of purchasing electricity to serve EV charging requests and the cost of battery degradation. The former is given by,
\begin{equation}
C^E= \sum_{t \in T} \rho_t \times p_t^E \times \Delta t,
\end{equation}
where $\rho_t$ and $p_t^E$ denote the energy price and charging power at time slot $t$, and we assume both remain constant during time duration $\Delta t$. The latter for battery degradation is based on \cite{chai2021optimal} and given by,
\begin{equation}
C^D= c^B \times \frac{r^D \times (1 + r^D)^{T^{\text{cycle}}}}{(1 + r^D)^{T^{\text{cycle}}} - 1},
\end{equation}
where $c^B$, $r^D$, and $T^{\text{cycle}}$ mean the cost for battery investment and discount rate, and the battery cycle life, respectively. Specifically, $T^{\text{cycle}}$ can be calculated by Eq. (\ref{eq-t-cycle}) as below.
\begin{equation}
\label{eq-t-cycle}
T^{\text{cycle}} = \frac{n^*}{365 \times 0.5 \times \sum_{t \in T} (\text{DoD}_t)^{\kappa}},
\end{equation}
where $n^*$ is the battery cycle life in years assuming 100\% DoD, which in reality may vary among different charging/discharging events. We let $\text{DoD}_t$ be the actual DoD for event $t$ and assume a set of events $T$ where $t\in T$. We also assume a uniformly 0.5 cycle for each charging/discharging event. Finally, $\kappa$ is a constant range from 0.8 to 2.1 empirically. Altogether, we present the first objective function $F_1$ as the total cost $C$ given by,
\begin{equation}
\label{eq-f1}
F_1:~C = C^E + \lambda_1 C^D,
\end{equation}
where the importance of battery degradation is controlled by a parameter $\lambda_1$.

\subsection{Objective 2: Distribution Network Performance}
We capture the performance of a distribution network by two important aspects, energy loss and voltage deviation. The former shall be minimized and the daily loss is given by,
\begin{equation}
    L =\sum_{t \in T}p^L_t,
\end{equation}
where $p^L_t$ represents the active power loss in distribution network in time $t$, within which $p^L_t$ remains unchanged. The latter shall be minimized as well and we model the daily voltage deviation as,
\begin{equation}
    V= \sum_{t \in T}\sum_{b\in B}\left|1-v_{t,b}\right|,
\end{equation}
where $B$ is the set of all buses and nodes and $v_{t,b}$ is the voltage for bus $b$ at time $t$. Altogether, we have the second objective function as the addition of both energy loss and voltage deviation, given by,
\begin{equation}
\label{eq-f2}
F_2:~L + \lambda_2 V,
\end{equation}
where $\lambda_2$ is a control parameter similar to $\lambda_1$. 

\subsection{Constraints}
\label{sec-prob-constraint}
The objective functions shall be optimized subject to several constraints, based on realistic settings for distribution networks. For buses and nodes, we need to maintain the balance of active and reactive power at every time slot $t$, as below.
\begin{equation}
\begin{split}
& p^A_t + \sum_{b\in B} p^{PV}_{t,b} + \sum_{b\in B} p^{BESS^-}_{t,b} \\= & \sum_{b\in B} p_{t,b}^R + \sum_{b\in B} p_{t,b}^L + \sum_{b\in B} p_{b,t}^{EV} + \sum_{b\in B} p_{t,b}^{BESS^+},
\end{split}
\end{equation}
where $p^A$ is the active power from the grid as one supply source of the network, $p^{PV}_{t,b}$ is the PV generated power to bus $b$, $p^{BESS^-}$ and $p^{BESS^+}$ are the discharging and charging power of BESS, respectively, and the EV charging load is captured by $p^{EV}$. Power can be non-positive, e.g., $p^{PV}_{t,b'}=0$ if no PVs are attached to bus $b'$. We also enforce a range of valid bus voltage as $[v^{\min}, v^{\max}]$, and the maximum apparent power is ${S_b}^{\max}$ for bus $b$. The exponential voltage-dependent active power demand is given by, 
\begin{equation}
p_{t,b}^A=p_{t,b}^{A^*}\left(\frac{v_{t,b}}{v_{t,b}^*}\right)^{\kappa_b^p}.
\end{equation}
where $v^*$ is the nominal voltage and $\kappa_b^p$ is a constant. Finally, we model the change of SoC of BESS at time $t$ as,
\begin{equation}
\text{SoC}_{t+1} = \text{SoC}_t - \frac{\Delta t}{\eta^- E^B} p_{t}^{BESS-} + \frac{\Delta t}{\eta^+ E^B} p_{t}^{BESS+},
\end{equation}
where $\eta^+$ and $\eta^-$ are the BESS efficiency for charging and discharging, respectively, and $E^B$ is the battery capacity in kWh. Besides, the valid range of BESS power is $[0, p^{BESS^*}]$.

\section{Methodology}
For our multi-objective problem, we present \texttt{Bach} to produce optimal schedules based on two different methods.

\subsection{$\varepsilon$-Constraint Method}
Given two objectives $F_1$ and $F_2$, as well as a set of constraints, we proposed to employ the $\varepsilon$-constraint method to perform multi-objective optimization and derive optimal solutions for \texttt{Bach}. The main idea of the method is to optimize one objective while treating the other one as an inequality constraint controlled by a parameter $\varepsilon$. Specifically, we let $F_1$ for monetary cost be the former and treat $F_2$ for distribution network performance as an inequality constraint, i.e.,
\begin{equation}
\label{eq-opt}
\begin{split}
    \min &~ F_1 \\
    \text{s.t.} &~ F_2\le\varepsilon \text{~and constraints in Section \ref{sec-prob-constraint}}
\end{split}
\end{equation}
where $\varepsilon$ systematically varies from its minimum to maximum for $F_2$ and the optimization problem in Eq. (\ref{eq-opt}) is solved for each sampled $\varepsilon$ value. Such an iterative method gradually adjusts $\varepsilon$ to identify a set of non-inferior solutions, which together form the Pareto front for our bi-objective optimization.

\subsection{Fuzzy Decision-Making Method}
In realistic scenarios, decision makers may prefer one optimal solution instead of many. The challenge for such one solution is to balance the multiple objectives simultaneously. We propose to employ a fuzzy decision-making method, which transforms each of the objectives into a fuzzy variable scaled between 0 and 1. The linear membership for such transformation for objective $i$ is given by, 
\begin{equation}
\label{eq-member-func}
\mu(F_i)=
\begin{cases}
1 & \text{if } F_i \leq F_i^{\min}   \\
\frac{F_i^{\max}-F_i}{F_i^{\max}-F_i^{\min}} & \text{if } F_i^{\min} < F_i < F_i^{\max} \\
0 & \text{if } F_i \geq  F_i^{\max}
\end{cases},
\end{equation}
where $F_i^{\min}$ and $F_i^{\max}$ represent the minimum and maximum  $F_i$ values, respectively. The membership value of 1 implies complete compatibility and 0 signifies incompatibility. Accordingly, we normalize the membership functions for each non-inferior solution, which can be calculated as,
\begin{equation}
    \mu^k=\frac{\sum_{i=1}^{m}{\mu(F_i^k)}}{\sum_{k=1}^{k}\big(\sum_{i=1}^{m}{\mu(F_i^k)\big)}},
\end{equation}
where $m=2$ is the number of objectives and $k$ is the number of solutions in this research.

\section{Experimental Study}
We present our experimental setup and results here.

\subsection{Experimental Setup}
\begin{figure}
    \centering
    \def \tmph{1.37in}
    \subfigure[PV and Price]{\includegraphics[height=\tmph]{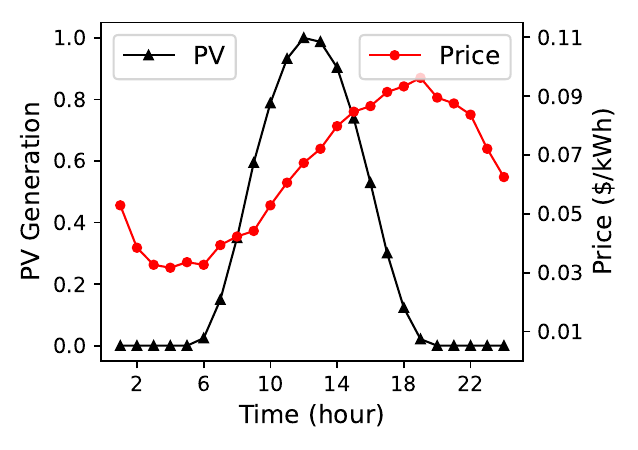}}
    \hspace{-1em}
    \subfigure[EV Load]{\includegraphics[height=\tmph]{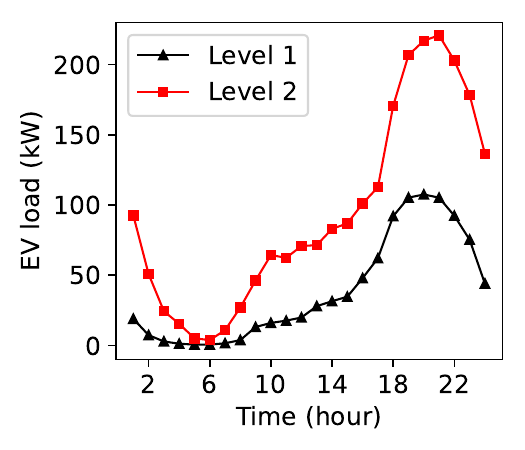}}
    \caption{Sample daily data for PV generation, electricity price, and EV load, with 150 EVs for each level.}
    \label{fig-input}
\end{figure}

Our multi-objective optimization methods and BESS scheduling in \texttt{Bach} utilize the GAMS solver and MATLAB interfacing \cite{babu2016optimal}, e.g., DICOPT from GAMS specialized for mixed-integer non-linear programming. We test \texttt{Bach} on the IEEE 33-bus distribution network with detailed parameters in Table \ref{tab-33bus}. Different customer types, e.g., commercial, residential, and industrial, along with their corresponding bus locations in the network are available in \cite{pamshetti2019optimal}. We take the load profile, price, PV generation, and EV load (levels 1 and 2) from \cite{pamshetti2019optimal,zhu2022producing} as the input of our methods, and a sample daily data is visualized in Fig. \ref{fig-input}. Besides, both $\lambda_1$ and $\lambda_2$ are equal to 1 by default. 

\begin{table}[]
\centering
\caption{Configurations of our 33-bus distribution network.}
\label{tab-33bus}
\renewcommand{\arraystretch}{1.3}
\begin{tabular}{ccc}
\hline\hline
 & Bus Identifiers & Capacity \\ \hline
PV & 9, 13, 25, 30 & 500 kW \\ \hline
EV & Level 1: 8, 14, 18; Level 2: 2, 19, 27 & 150 EVs \\ \hline
BESS & 18, 33 & 1 MWh \\ \hline\hline
\end{tabular}
\end{table}

\subsection{Results and Discussions}
To demonstrate the effectiveness of our multi-objective optimization, we implement and discuss the following cases.

\begin{itemize}
    \item Case 1: Optimize $F_1$ only for minimizing the costs for electricity purchase and BESS degradation.
    \item Case 2: Optimize $F_2$ only for maximizing the network performance in terms of power loss and voltage deviation.
    \item Case 3: Optimize both $F_1$ and $F_2$ for optimizing both monetary cost and network performance.
\end{itemize}

We present the results of the three cases in Table \ref{tab-results} for energy cost, degradation cost, energy loss, and voltage deviation, and detailed results discussions below.

\begin{table}[!t]
\centering
\renewcommand{\arraystretch}{1.3}
\caption{The results of \texttt{Bach} for different cases, where Cases 1 and 2 have minimum cost and network performance, respectively, and Case 3 achieve the optimal for both cost and performance.}
\label{tab-results}
\begin{tabular}{ccrrr}
\hline\hline
\multicolumn{2}{c}{Objective Function} & Case 1 & Case 2 & Case 3 \\ \hline
\multirow{2}{*}{$F_1$} & energy purchase (\$) & 3,860.18 & 3,879.01 & \textbf{3,851.59} \\ \cline{2-5} & battery degradation (\$) & \textbf{142.58}  & 499.26  & 218.09  \\ \hline
\multirow{2}{*}{$F_2$} & energy losses (kWh) & 1,520.36 & \textbf{1,206.44} & 1,276.78 \\ \cline{2-5} & voltage deviations & 19.16 & \textbf{15.75} & 15.92   \\ \hline\hline
\end{tabular}
\end{table}

\subsubsection{Case 1}
This case is cost driven. The cost, for both energy purchase and battery degradation, is naturally low, evidenced by the results in Table \ref{tab-results}. Specifically, the energy expenditure is 3,860, which is lower than the cost of Case 2. The degradation cost is 143, which is only 29\% and 66\% of the costs of Case 2 and Case 3, respectively. The performance of distribution networks is not within the consideration of this case, and the values of energy losses and voltage deviation are high, e.g., 26\% and 22\% higher compared to Case 2 for losses and deviation, respectively. Note that minimizing the cost without considering the network performance is often not realistic, which sacrifices or even violates the service quality.

\subsubsection{Case 2}
Different from Case 1, we optimize the network performance only in this case with the cost non-optimized. Table \ref{tab-results} shows that this case has the best network performance. The energy loss is 1,206 kWh, nearly 20\% less compared to the above case. The voltage deviation is 15.75, 22\% lower than the cost-driven case. Without focusing on the cost, this case incurs higher costs for energy purchase and battery degradation compared to Case 1; especially for degradation, the cost is 3.5x higher. While the network performance is competitive, this case may not be competitive for a long-term business. 

\subsubsection{Case 3}
\begin{figure}
    \centering
    \def \tmph{1.45in}
    \subfigure[$F_1$ vs. $F_2$]{\includegraphics[height=\tmph]{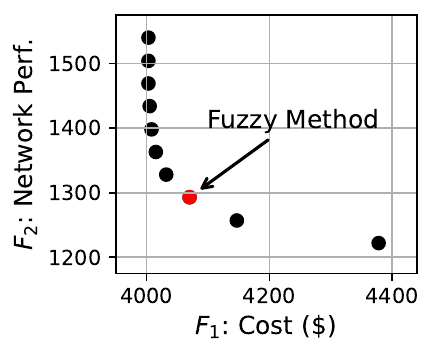}\label{fig-pareto-a}}
    \hspace{-1em}
    \subfigure[Battery Life vs. $F_2$]{\includegraphics[height=\tmph]{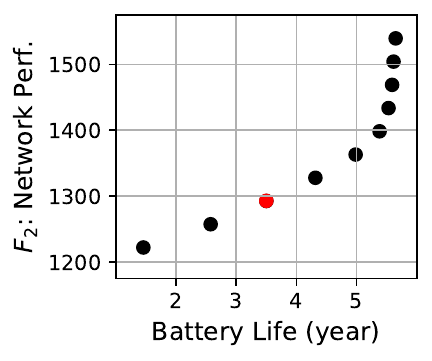}\label{fig-pareto-b}}
    \caption{Case 3 with multiple optimal solutions for both objectives for cost minimization and network performance as a Pareto front. The solution based on our fuzzy method for decision-making is annotated and highlighted in red.}
\label{fig-pareto}
\end{figure}

This case optimizes both objectives and models a more realistic scenario. Minimizing the cost and maximizing the network performance is sometimes contradictory, meaning the improvement of one objective comes with the decreased performance of the other. Some solutions do not dominate each other in both objectives and multiple optimal solutions exist with different emphases of the objectives. We show the solutions in Fig. \ref{fig-pareto-a}, where $F_1$ for this case ranges from \$4,003 to \$4,378 and $F_2$ ranges from 1,222 to 1,540. Among these solutions, our proposed fuzzy method with linear membership identifies one of the solutions for decision-making, which is highlighted in Fig. \ref{fig-pareto}. Specifically for this solution, the costs for energy purchase and battery degradation are \$3,852 and \$218, respectively, while the energy loss and voltage deviation are 1,277 kWh and 15.9, respectively. We also show the results for battery life in Fig. \ref{fig-pareto-b} where we can observe that battery life needs to be compromised for optimizing the distribution network. Overall, the results of this case are competitive compared to the other two cases as shown in Table \ref{tab-results}, and show a great balance in optimizing both monetary cost and network performance.

\begin{figure}
    \centering
    \def \tmph{1.415in}
    \subfigure[Variation of $\lambda_1$]{\includegraphics[height=\tmph]{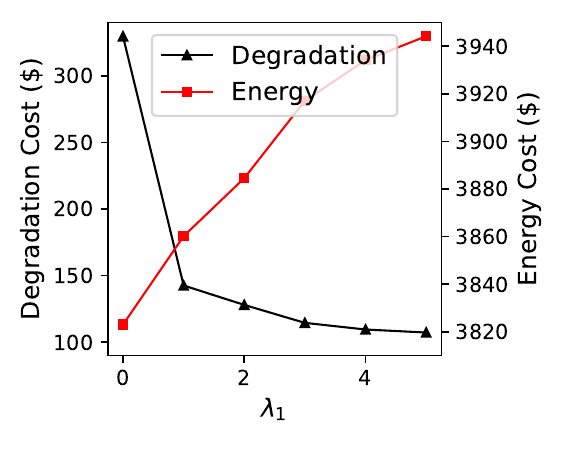}\label{fig-lambda1}}
    \hspace{-1em}
    \subfigure[Variation of $\lambda_2$]{\includegraphics[height=\tmph]{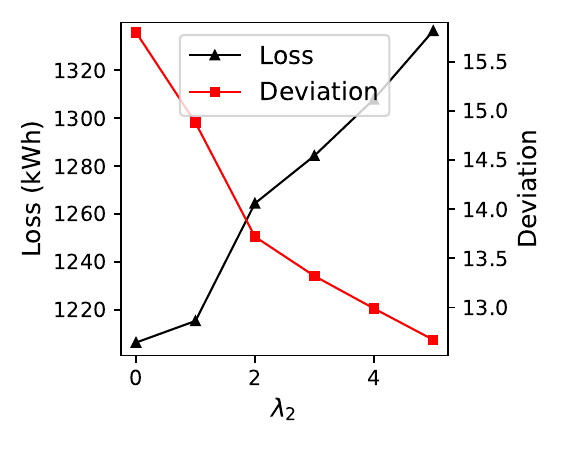}\label{fig-lambda2}}
    \caption{The parameter sensitivity analysis results with varying $\lambda_1$ for the cost objective and $\lambda_2$ for the network performance objective. Large $\lambda_1$ emphasizes the importance of battery degradation and voltage deviation becomes more important with large $\lambda_2$.}
    \label{fig-lambda}
\end{figure}

\subsection{Parameter Sensitivity Analysis}
We investigate the performance sensitivity of two parameters, $\lambda_1$ and $\lambda_2$, for controlling the importance of different terms of the objectives. The results for $F_1$ are illustrated in Fig. \ref{fig-lambda1}. We can see that the cost of battery degradation decreases with large $\lambda_1$, and the observation aligns with our $F_1$ settings in Eq. (\ref{eq-f1}) where a large $\lambda_1$ increases the importance of battery degradation. The emphasis of one implies the reduced importance of the other, and we can observe that the energy cost goes up when the degradation cost lowers. This parameter is helpful for practical usage of \texttt{Bach} with customised configurations for different applications. Furthermore, the importance of energy loss and voltage deviation varies in different scenarios in reality, e.g., mission critical systems like data centers have stringent requirements for voltage deviation. Our formulation supports customized control of both factors by adjusting $\lambda_2$ in $F_2$, and we show our results in Fig. \ref{fig-lambda2}. With large $\lambda_2$, voltage deviation has a big weight in $F_2$ and there is a downtrend (means improvement) of deviation as $\lambda_2$ increases. However, this comes with a cost, i.e., an uptrend of energy loss. 

\section{Conclusion}
In this paper, we present \texttt{Bach}, for battery degradation-aware scheduling with a novel multi-objective framework. \texttt{Bach} considers both monetary cost and network performance for optimization, where the former is based on energy expenditure and battery degradation and the latter incorporates energy loss and voltage deviation. The experiments, based on the IEEE 33-bus network, demonstrate \texttt{Bach}'s effectiveness in saving costs with extended battery life and maintaining competitive network performance. \texttt{Bach}, therefore, has the potential to contribute to sustainable and resilient energy systems.

\balance
\bibliographystyle{IEEEtran}
\bibliography{IEEEabrv,referencestencon}

\end{document}